# Data Driven Prediction of Battery Cycle Life Before Capacity Degradation

Team 4

Caitlin Feltner (cmf269); Kurt I. Kuhn (kik35); Jamie Peck (jp2484); Anmol Singh (as2753)

SYSEN 5880 Project Proposal

May 20, 2020

## Abstract

Ubiquitous use of lithium-ion batteries across multiple industries presents an opportunity to explore cost saving initiatives as the price to performance ratio continually decreases in a competitive environment. Manufacturers using lithium-ion batteries ranging in applications from mobile phones to electric vehicles need to know how long batteries will last for a given service life. To understand this, expensive testing is required.

This paper utilizes the data and methods implemented by Kristen A. Severson, et al, to explore the methodologies that the research team used and presents another method to compare predicted results vs. actual test data for battery capacity fade. The fundamental effort is to find out if machine learning techniques may be trained to use early life cycle data in order to accurately predict battery capacity over the battery life cycle. Results show comparison of methods between Gaussian Process Regression (GPR) and Elastic Net Regression (ENR) and highlight key data features used from the extensive dataset found in the work of Severson, et al.

# 1. Introduction

Lithium-ion batteries are one of the most widely used batteries today mainly because of their continual dropping price in industry, high energy densities, high cell working voltage, and good power capability. They comprise a large percentage of the batteries used in popular everyday electronics such as cell phones, laptops, and gaming systems, for example. They are also becoming more prominent in the military and aerospace industries and are dominating in the transportation industry sector, specifically for electric vehicles.[1] That being stated, companies and consumers need to know the cycle lifetime of the battery, defined as the number of charging and discharging cycles after which the battery capacity drops below 80% of the nominal value, primarily for commercial and technical reasons relating to warranty and remaining useful life of the product overall.[2][7]

Physically testing the complete cycle life can take, on average, 5 hours for one cycle[10] and accurately predicting the nonlinear life of a lithium-ion battery through machine learning modeling methods will allow the performance of the chemistry relative to the loading conditions to be predicted earlier and with minimal resources and cost. These modeling techniques are additionally attractive right now for battery management systems (BMS) and enhances their ability to perform BMS functions, including: controlling, calculating, and monitoring the state-of-function in the form of state-of-charge and state-of-health; protecting the battery system for longevity; managing the functional safety aspects of the battery system relating to high temperature, cell imbalance, or calibration of system; and, indicating end of life when capacity of the battery system falls below the user-set target threshold.[8] Using machine learning predictive modeling techniques is a challenge for this application because of the nonlinear degradation process, as well as the limited datasets used to date that do not span across the range of life, but the drive from industry makes this an area that needs to be further researched.[2]

In this study, the team hypothesizes that a machine learning model can be developed in order to accelerate the development of lithium-ion battery technology and meet the industry demand for creating a data-driven model that accurately predicts the cycle life of commercial lithium iron phosphate(LFP)/graphite cells using early-cycle data. Because the graphite negative electrode is the most significant driving force for degradation in these cells, this model and the results will be useful for commercial lithium-ion battery chemistries that are based on graphite.[2]

## 2. Background

Since lithium-ion batteries have a long service lifetime, feedback on performance is delayed by months or years. Due to this delayed feedback, research in previous studies has sought to accurately estimate battery life through mechanistic and semi-empirical computer modeling. However, predicting the performance of lithium-ion batteries is challenging due to nonlinear degradation associated with battery charging cycles. This study uses the work from Severson, et al. to show the use of machine learning to accurately predict lithium-ion capacity degradation from early-cycle discharge data.[2]

Prior work cited by the study[2] shows predictive models used "diverse mechanisms such as growth of the solid–electrolyte interphase, lithium plating, active material loss, and impedance increase" in order to determine remaining useful life. In addition, "coulombic efficiency and impedance spectroscopy" are cited as specialized diagnostic measurements to determine useful lifetime estimates. These chemistry and mechanistic approaches showed predictive success; however, Severson, et al. sought to use a dataset that considers relevant operating conditions (such as fast charging) as a mechanism-agnostic alternative to prior studies.

Predicting the performance is difficult because of the non-linear nature of the degradation process shown in *Figure 1*. This figure shows LFP/graphite cell discharge capacity (Ah) vs. cycle number for a data set associated with the work of Severson, et al. What is interesting is that the crossing of the capacity fade over the number of cycles shows a weak correlation to the initial cell capacity. Therefore, high rate discharge voltage curves are used as opposed to only capacity fade curves.

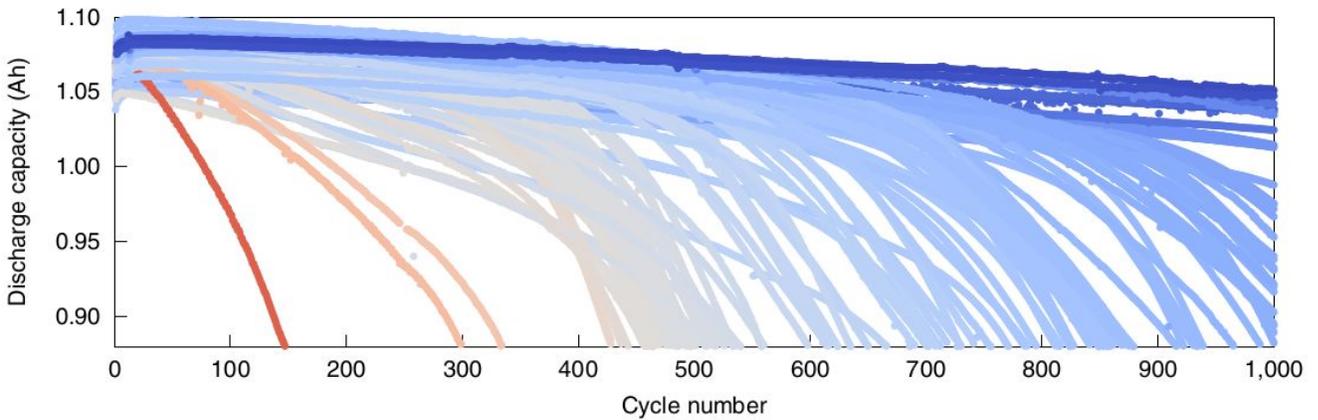

*Figure 1: Discharge Capacity vs. Cycle Number[2]*

Data generated by Severson, et al. consists of 124 cells cycled between 150 - 2300 times using 72 fast-charging conditions for a total dataset of approximately 96,700 cycles. Varying the charging conditions was a key contributor to the wide range of cycle lives shown in *Figure 1*.

Charge, discharge, and depth of discharge (DOD) under different loading conditions and temperatures all contribute to a reduction in battery life. For this study, A123 Systems cells, model APR18650M1A, 1.1Ah nominal capacity, were cycled in a temperature-controlled environmental chamber at 30 degrees celsius under various fast-charging conditions but with identical discharging conditions of 4C to 2.0V, where 1C is 1.1A, to create the initial dataset. Data from 124 cells with cycle lives ranging from 150 to 2,300 using 72 different fast-charging conditions were collected, with end of life defined as the number of cycles until 80% of nominal capacity. Within these 72 different fast-charging conditions, charging rates ranged from 3.6C, which was the manufacturer's recommended fast-charging rate, to extreme fast-charging (XFC) conditions of 6C[2]. XFC conditions, defined at around 10 minutes or less for power cells like the one studied here from A123 Systems, is an area being driven by industry demand and worthwhile to study further[9]. By varying charging conditions, this creates a dataset that has a wide range of cycle lives *(Figure 1)*, averaging cycle life around 806 with a standard deviation of 377 ranging from approximately 150 cycles to 2,300 cycles. Temperature controlling these different fast-charging profiles varied by around 10 degrees celsius within the cycle due to the large amount of heat generation during charge and discharge for this dataset[2].

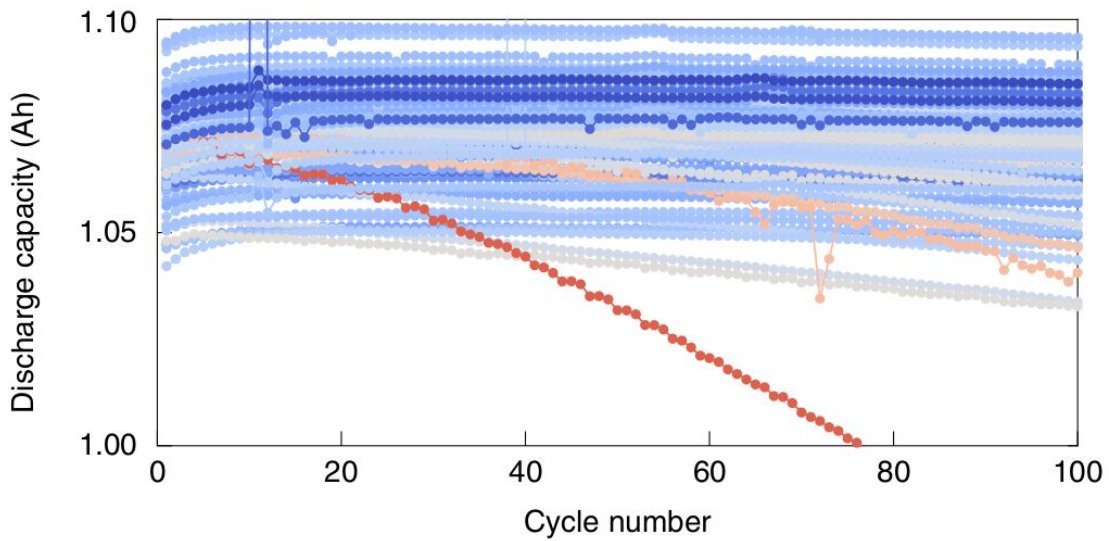

*Figure 2: Discharge Capacity vs. Cycle Number for First 100 Cycles[2]*

The voltage curve is a particularly important parameter when studying the cycle life of lithium-ion batteries because it can tell us a lot about a battery's degradation diagnosis[2]. In particular, the physical variance parameter is associated with the amount of discharged energy dissipation on voltage. Furthermore, the integral of the variance parameter is associated with the change in energy dissipation between cycles. Therefore, it can be stated that the "variance of ΔQ(V) reflects the extent of non-uniformity in the energy dissipation with voltage…"[2] and this is why it is important to the study.. As shown in *Figure 3* below, the less the variance, the better the cycle life of the battery. Also, the high correlation between variance of ΔQ(V) and battery life cycle makes this useful for a machine learning approach to predicting battery life.

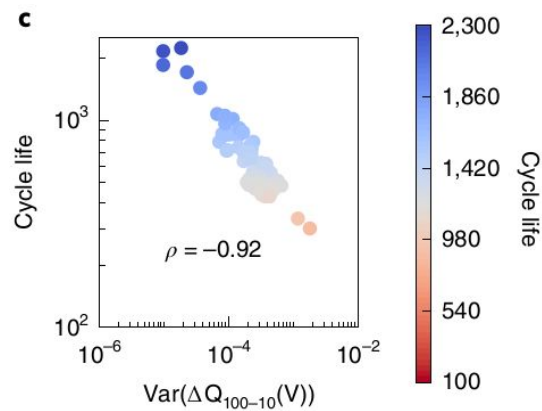

*Figure 3: Cycle life vs. Var(ΔQ$_{100-10}$(V))[2]*

State of Charge (SOC), useful capacity and power capability are not measurable and must be estimated using battery modelling. Moreover, the behavior of the battery and its internal state changes over time and usage as the battery degrades further complicating modelling of the battery as the estimates of SOC; useful capacity and power capability must be available through the entire lifetime of the battery[5]. The State of Health (SOH) of a battery reflects two phenomena that occur as batteries age: progressive loss of storage capacity, which is known as capacity fade (an irreversible loss of the ability of a battery to store charge); and, progressive increase of impedance (an irreversible reduction of the rate at which electrical energy can be accepted or released by the battery), which causes the power provided by the battery to decline. Estimates of a battery's SOH, therefore, have to take into consideration both capacity fade and impedance increase. Being complex systems, capacity fade and increase in impedance of Li-ion batteries occurs due to several physical and chemical processes as seen in *Figure 4*. These mechanisms depend on many factors such as battery chemistry, fabrication, operating conditions and usage history[3]. The degradation mechanisms can be better understood by classifying them as-

- Calendar fade- SEI growth, loss of cyclable Li coupled with cycling
- Cycling fade- Active material structure degradation, mechanical fracture

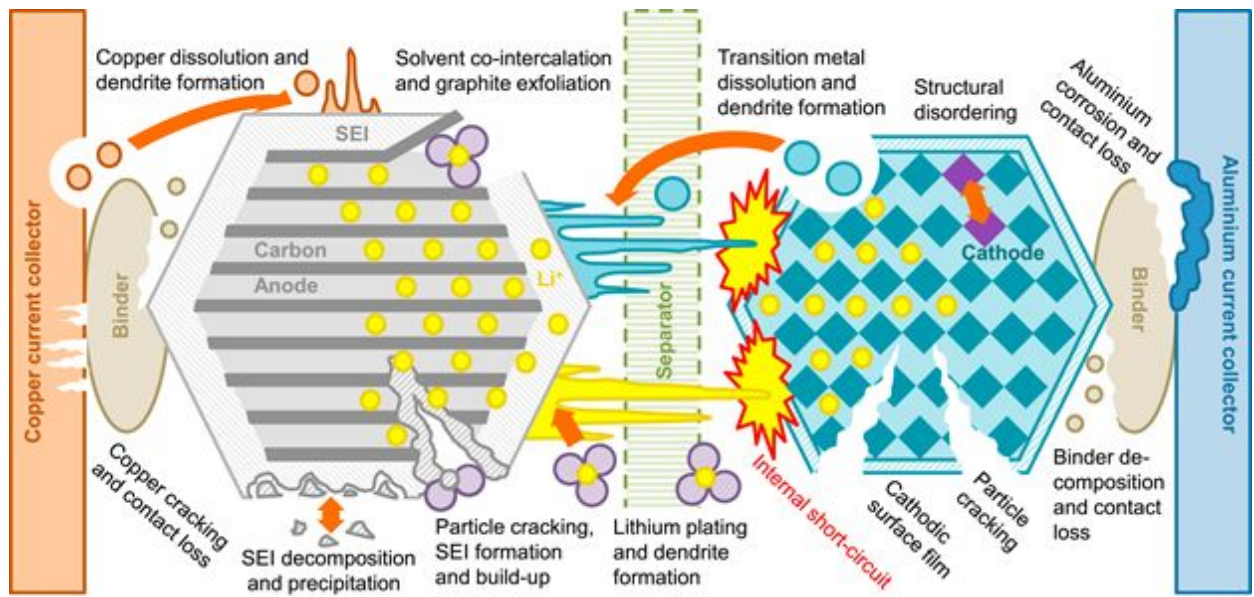

*Figure 4: Degradation mechanisms in Li-ion cells[4]*

While chemistry and/or mechanism-specific models[11-12] have shown predictive success, developing models that describe full cells cycled under relevant operating conditions (for example, fast charging) remains challenging, given the many degradation modes and their coupling to thermal and mechanical heterogeneities within a cell. However, approaches using statistical and machine-learning techniques to predict cycle life are attractive, mechanism-agnostic alternatives and assume no prior knowledge of cell chemistry and degradation mechanisms thus providing opportunities for higher accuracy, earlier prediction, greater interpretability and broader application to a wide range of cycling conditions.

# 3. Methods

Referencing the available dataset of Severson, et al., the team focused on developing two different machine learning models and ultimately determined which model provided the most accurate prediction of cycle life using the first 100 beginning of life cycles where capacity fade is not yet exhibited in the electrochemistry. The two selected models that were compared are the Gaussian Process Regression model and the Elastic Net Regression model. Later in the Results and Discussions section of this report, information on each model details and assumptions will be provided.

The leveraged dataset used approximately the first 100 cycles of 124 lithium-iron phosphate (LFP)/graphite cells. As indicated earlier, this is a significant dataset because graphite is commonly a large contributor to capacity fade at end of life conditions, therefore parallels can be drawn between LFP/graphite cells and common Li-ion cells available on the market today[4]. 41 cells make up Batch 1, 43 cells make up Batch 2, and 40 cells make up Batch 3. For our analysis, all three batches were combined to develop the two models built in order to improve accuracy. Within the dataset, there are time-scalar features that were collected, including: current, charge capacity, discharge capacity, various temperature metrics, and voltage, for example. These time scalar features were collected over the course of all cycles to end of life failure, defined previously in the Background section. Cells were charged and discharged over varied charge and discharge loading conditions, some more abusive than others. For example, extreme fast charging conditions where cells were charged at 6C rate, which will cause cells to reach end of life sooner, were compared to cells that were charged at a 3.6C rate, which will allow cells to complete more cycles over life and reach end of life slower, comparatively. Within the dataset there were scalar features recorded as well, including: internal resistance (DCIR), total charge capacity, total discharge capacity, total time, and various temperature statistics. The same dataset processing methods were used as Severson, et. al., which removes outliers of cells that had early cycle life failures and additionally combined all 3 batches of cells for model build and analysis. Overall, greater than 80,000 data points were used for analysis.

Seven different features were considered for dataset training and modeling. These features included minimum change in discharge capacity from cycles 100 to 10, variance of discharge capacity difference between cycles 100 and 10, internal resistance difference between cycle 2 and 100, mean of the charge time of cycles 2 to 6, max temperature from cycle 2 to 100, slope of discharge curve, and integral of average temperature over cycles 2 to 100. These features were selected based on industry experiences and previous work from Severson, et. al.

Below, *Equation 1* defines the change in discharge voltage curves as a function of voltage from cycles 100 to 2, and *Equation 2* defines the mean of the change in discharge voltage curves as a function of voltage from a predefined cycle count, p. These values are used commonly in the features selected, so they are defined separately from the features considered for the models.

*Equation 1:*

$$\Delta Q(V) = Q_{100}(V) - Q_{10}(V), \; \Delta Q(V) \, \varepsilon R^p$$

*Equation 2:*

$$\overline{\Delta Q(V)} = \frac{1}{p} \sum_{i=1}^{p} \Delta Q(V)$$

DelQ is the change in discharge voltage curve as a function of voltage for a given cycle. This can be used because voltage range is identical for every cycle, therefore capacity can be considered as a function of voltage to maintain a uniform basis for comparing cycles.

*Equation 3* and *Equation 4* focus on the change in discharge voltage curves as a function of voltage, where *Equation 3* is the minimum delQ for 100 to 10 cycles and *Equation 4* is the variance of delQ for 100 to 10 cycles. When log is referenced in all equations, it implies log base 10, and again p is in reference to cycle number.

*Equation 3:*

$$Minimum = log(|min(\Delta Q(V))|)$$

*Equation 4:*

$$Variance = \log\left(\left|\frac{1}{p-1}\sum_{i=1}^{p}(\Delta Q(V) - \overline{\Delta Q(V)})^2\right|\right)$$

The only discharge capacity fade curve feature considered within the models is where the slope of the linear fit to the capacity fade curve from cycles 2 to 100 was considered. The slope of the discharge curve from cycles 2 to 100 is calculated as the first value in vector $b^*$, shown in *Equation 5*, where d, the number of cycles used in the prediction, are equal to 99. For *Equation 5*, q $\varepsilon R^d$ is the vector of discharge capacities as a function of the cycle number, X $\varepsilon R^{dx2}$ is the cycle numbers used in prediction within the first column and the second column are ones, and b $\varepsilon R^2$ is a coefficient vector. *Equation 6* shows the discharge capacity fade curve feature in equation form.

*Equation 5:*

$$b^* = arg_b min \frac{1}{d}\|q - Xb\|_2^2$$

*Equation 6:*

*Slope of Discharge Curve, cycles* 2 *to* 100 $=$ *first value in vector* $b^*$ *where d* $=$ 99

*Equation 7, Equation 8, Equation 9* and *Equation 10* all focus on other features that impact cell capacity fade over life. In *Equation 7,* the mean charge time for the first five cycles were considered. In *Equation 8*, the max temperature was considered from cycle 2 to 100. In *Equation 9*, the internal resistance between cycles 100 and 2 were considered. For *Equation 10*, the integral of temperature over time for cycles 2 to 100 were considered.

*Equation 7:*

$$Average\ Charge\ Time = \frac{1}{5}\sum_{i=2}^{6} Charge\ Time_i$$

*Equation 8:*

*Maximum Temperature, cycles* 2 *to* 100 $= max_n T(n)$

*Equation 9:*

$$\text{Internal Resistance, cycle } 100 - \text{cycle } 2 = IR(n=100) - IR(n=2)$$

*Equation 10:*

$$\text{Temperature Integral, cycle 2 to 100} = \int_{t_2}^{t_{100}} T(t)dt$$

Prediction accuracy is determined by calculating the root-mean-squared error (RMSE) in cycles and average percentage error[2]. Below, *Equation 11* and *Equation 12* describe RMSE and average percentage error, respectively:

*Equation 11:*

$$RMSE = \sqrt{\frac{1}{n}\sum_{i=1}^{n}(y_i - \widehat{y_i})^2}$$

*Equation 12:*

$$\% \, err = \frac{1}{n}\sum_{i=1}^{n}\frac{|y_i - \widehat{y_i}|}{y_i}x100$$

The model with the smallest percentage error and RMSE was deemed the preferred method for accuracy, and conclusions were drawn based on these methods used to generate these results. The models are deemed successful if accuracy is relatively low and comparable to other methods. From there, final conclusions about the true benefit of using either model and other machine-learning techniques to quantitatively predict life cycle performance before cell capacity degradation were made, and a final recommendation is referenced later in the document.

# 4. Results and Discussion

We present two models developed in MATLAB to predict battery cycle life using the seven features mentioned in the methods section. The training data are used to learn the model structure and coefficient values. The testing data are used to assess generalizability of the model.

## 4.1. Model 1: Gaussian Process Regression (GPR)

Model 1 was trained using GPR with kernel function 'ARDExponential' and initial parameters $\beta$= 1000 and $\sigma$= 100. GPR is a non-parametric kernel based probabilistic method ideal for modelling arbitrarily complex systems. It is capable of expressing uncertainty thus giving a credible interval consisting of probabilistic upper and lower bounds along with the results. ARDExponential is the built-in kernel function with a separate length scale for each predictor available in MATLAB. Length scale defines how far apart the input values $X_i$ can be for response values to become uncorrelated making it an appropriate choice for our non-linear dataset. GPR is formulated as:

$$f(\mathbf{x}) \sim \mathscr{GP}(m(\mathbf{x}), \kappa(\mathbf{x}, \mathbf{x}'))$$

where m(x) and κ(x, x′) are the mean and kernel functions respectively, denoted by

$$m(\mathbf{x}) = \mathbb{E}[f(\mathbf{x})]$$
$$\kappa(\mathbf{x}, \mathbf{x}') = \mathbb{E}[(f(\mathbf{x}) - m(\mathbf{x}))(f(\mathbf{x}') - m(\mathbf{x}'))^T]$$

For any finite collection of input points, say X = $x_1$, …,$x_{ND}$, this process defines a probability distribution p(f ($x_1$),…,f ($x_{ND}$)) that is jointly Gaussian, with some mean m(x) and covariance K(x) given by $K_{ij}$=κ($x_i$, $x_j$). In this work, we use the ARDExponential kernel function[15]:

$$k(x_i, x_j | \theta) = \sigma_f^2 \exp(-r),$$

where

$$r = \sqrt{\sum_{m=1}^{d} \frac{(x_{im} - x_{jm})^2}{\sigma_m^2}}.$$

The normalized Predictor weights shown in *Figure 5* gives us an insight into the relative importance given to the features by the GPR model while training the dataset. GPR has given more importance to min delQ, variance delQ and Slope of the linear fit to the capacity fade curve from cycles 2 to 100 compared to other selected features and has ignored Integral of temperature over time from cycles 2 to 100, while training the model.

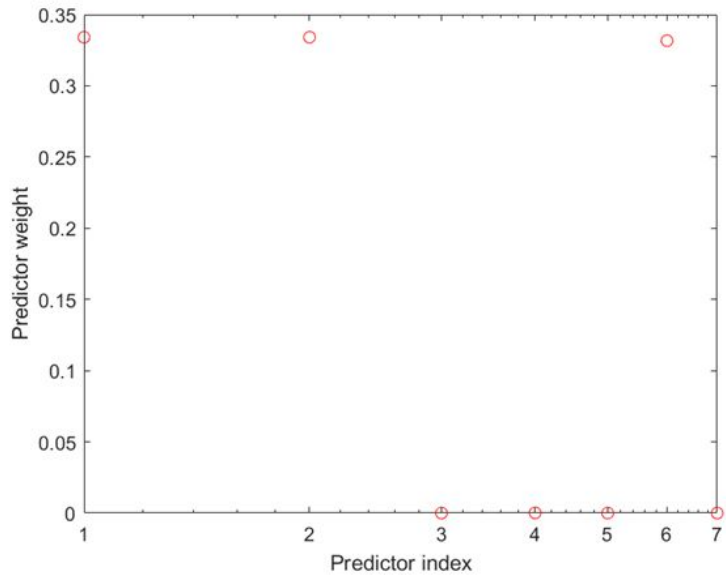

*Figure 5: Predictor weight vs. predictor index*

We can observe from *Figure 6* that GPR results in a well-fit model with observed cycle life falling within the standard deviation of the predicted cycle life. It has been successful in predicting the cycle life for batteries with low cycle life (<500 cycles) as well as for batteries with high cycle life (>1500) thus showing consistent results.

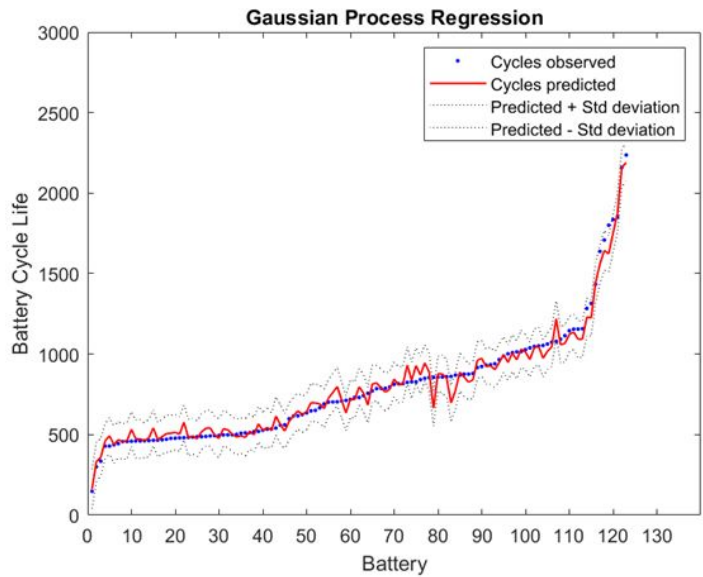

*Figure 6: Battery Cycle - Life vs. Count for GPR*

## 4.2. Model 2: Elastic Net Regression (ENR)

This model was used by Dr. Severson and her team to train the dataset. ENR is a regularized linear regression and it was used due to its low computation cost and ability to be trained offline. Model 2 was trained using ENR to perform model fitting and selection by finding sparse coefficient vectors. ENR is a special form of Lasso Regression with **α**=0.5. We applied four-fold cross validation as done in Severson et al., for our model development. The formulation for Elastic Net Regression is:

$$\hat{w} = \operatorname{argmin}_w \|y - Xw\|_2^2 + \lambda P(w)$$

$$P(w) = \frac{1-\alpha}{2} \|w\|_2^2 + \alpha \|w\|_1$$

The results of the cross validation was used to estimate the lambda with minimum Mean Square Error (**λ**= 0.0438) as depicted in *Figure 7* which helped determine the intercept and coefficients of the model.

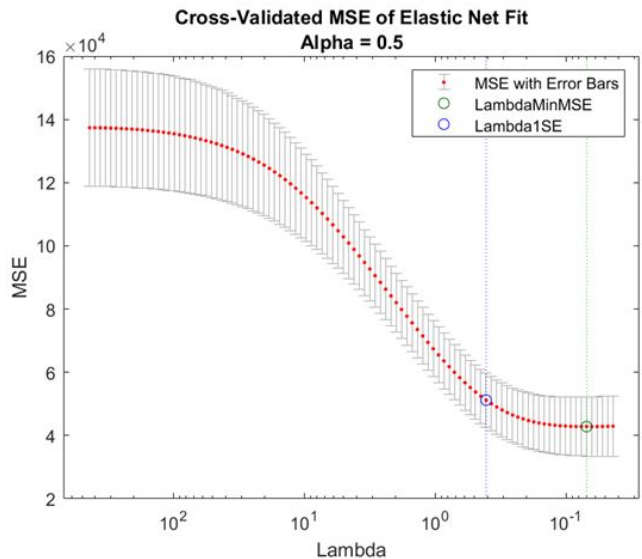

*Figure 7: Mean Squared Error (MSE) vs. Lambda*

The normalized Predictor coefficients illustrated in *Figure 8* gives us an insight into the relative importance given to the features by the ENR model while training the dataset. ENR has given more importance to Slope of the linear fit to the capacity fade curve from cycles 2 to 100 compared to other selected features and has ignored min del and variance delQ while training the model. Thus, we can confirm that both the trained models give significance to the Slope of the linear fit to the capacity fade curve from cycles 2 to 100. However, their considerations of other features are widely contrasting.

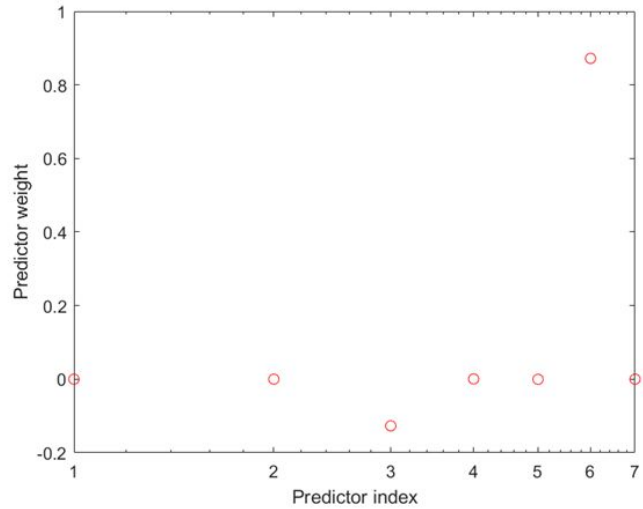

*Figure 8: Predictor weight vs. predictor index*

We can observe from *Figure 9* that ENR results in a model that fails to accurately predict the battery cycle life. It has been particularly unsuccessful in predicting the cycle life for batteries with high cycle life (>1500).

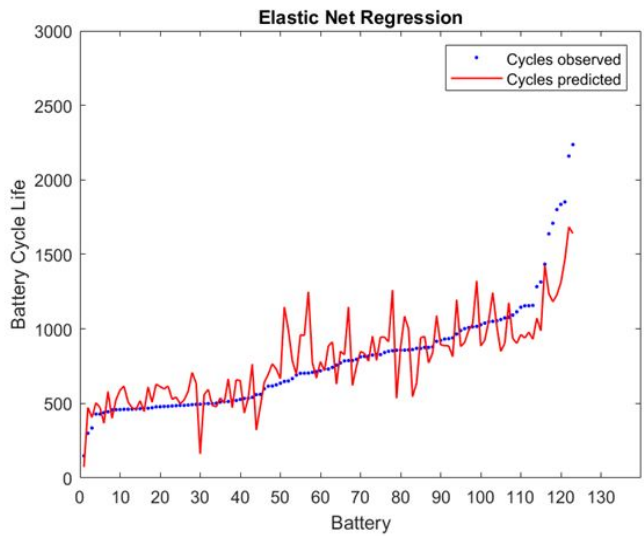

*Figure 9: Battery Cycle - Life vs. Count for ENR*

## 4.3. Comparative Analysis

*Table 1* and *Figures 10 & 11* display the performance of the GPR and ENR models applied to the dataset. The vertical dotted line indicates when the prediction is made in relation to the observed cycle life. The inset shows the histogram of residuals (predicted − observed) for the test data for the two models.

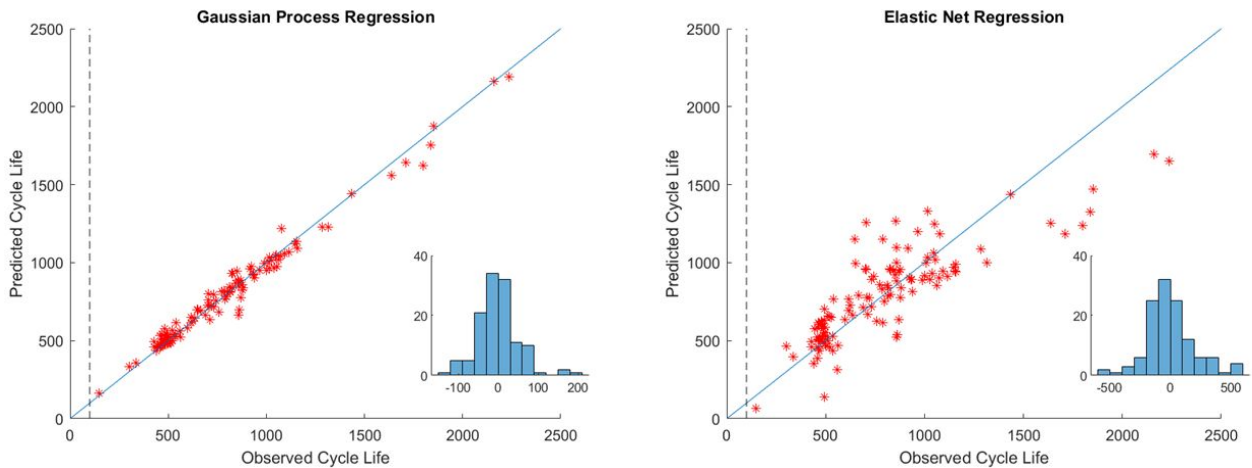

*Figures 10 & 11: Predicted vs. Observed Cycle Life for GSP & ENR*

The first model, using Gaussian Process Regression results in a model with approximately 5% average percentage error on the training dataset with root mean square error of 52 cycles. The second model, using Elastic Net Regression, is the model that is used in the research paper that we have used as our reference. Using the same features that we utilized to develop our first model, we can draw a comparison between the two regression methods for predicting battery cycle life testing the same dataset. The error average error percentage of the training dataset for the second model was calculated to be approximately 18% with root mean square error of 197 cycles.

*Table 1: Model metrics for the results shown in Figures 10 & 11*

| Model | RMSE (cycles) | Mean Percent Error (%) |
| --- | --- | --- |
| Gaussian Process Regression | 52 | 5 |
| Elastic Net Regression | 197 | 18 |

# 5. Conclusion and Recommendations

Based on the results of this study, it has been concluded that given the particular dataset and the specific features utilized from the dataset, Gaussian Process Regression is a better method to use than Elastic Net Regression when creating a model that will predict the cycle lifetime of a battery, defined as the number of charging and discharging cycles after which the battery capacity drops below 80% of the nominal value. This conclusion can be drawn based on the fact that the Gaussian Process Regression model provided a better fit with observed cycle life falling within the standard deviation of the predicted cycle life, and the Elastic Net Regression model was not able to accurately predict the cycle life for batteries with cycle lives greater than 1500 cycles.

There is potential for further extensions of this work. If the scope of the project were to be expanded and the time given to complete the study is not limited, the number of methods used to predict the cycle life of the lithium ion batteries can be expounded upon to included a number of methods such as k-means, SVM, naive bayes, etc. A more thorough comparison of different models would provide better insight into which model is most favorable for predicting the battery cycle life after which the battery capacity drops below 80%.

Finally, with added time a more thorough study between different Gaussian Process Regression kernels can also be explored. There are many kernels that could be used, and although we found one that provided favorable results, an analysis and comparison could be carried out to determine if there are Gaussian Process Regression kernels that would provide an even smaller mean percent error, i.e. less than 5%.

# Appendix

```matlab
load 2017-05-12_batchdata_updated_struct_errorcorrect.mat

batch1 = batch;
numBat1 = size(batch1,2);

load 2017-06-30_batchdata_updated_struct_errorcorrect.mat

%Some batteries continued from the first run into the second. We append
%those to the first batch before continuing.
add_len = [661, 980, 1059, 207, 481];
summary_var_list = {'cycle','QDischarge','QCharge','IR','Tmax','Tavg',...
    'Tmin','chargetime'};
batch2_idx = [8:10,16:17];
```

```matlab
for i=1:5
    batch1(i).cycles(end+1:end+add_len(i)+1) = batch(batch2_idx(i)).cycles;
    batch1(i).summary.cycle(end+1:end+add_len(i)+1) = ...
        batch(batch2_idx(i)).summary.cycle;
    batch1(i).summary.QDischarge(end+1:end+add_len(i)+1) = ...
        batch(batch2_idx(i)).summary.QDischarge;
    batch1(i).summary.QCharge(end+1:end+add_len(i)+1) = ...
        batch(batch2_idx(i)).summary.QCharge;
    batch1(i).summary.IR(end+1:end+add_len(i)+1) = ...
        batch(batch2_idx(i)).summary.IR;
    batch1(i).summary.Tmax(end+1:end+add_len(i)+1) = ...
        batch(batch2_idx(i)).summary.Tmax;
    batch1(i).summary.Tavg(end+1:end+add_len(i)+1) = ...
        batch(batch2_idx(i)).summary.Tavg;
    batch1(i).summary.Tmin(end+1:end+add_len(i)+1) = ...
        batch(batch2_idx(i)).summary.Tmin;
    batch1(i).summary.chargetime(end+1:end+add_len(i)+1) = ...
        batch(batch2_idx(i)).summary.chargetime;
end

batch([8:10,16:17]) = [];
batch2 = batch;
numBat2 = size(batch2,2);
clearvars batch
```

```matlab
load 2018-04-12_batchdata_updated_struct_errorcorrect.mat
batch3 = batch;
```

```matlab
batch3(38) = []; %remove channel 46 upfront; there was a problem with
%the data collection for this channel
numBat3 = size(batch3,2);
endcap3 = zeros(numBat3,1);
clearvars batch
for i = 1:numBat3
    endcap3(i) = batch3(i).summary.QDischarge(end);
```

```matlab
    end

rind = find(endcap3 > 0.885);
batch3(rind) = [];
```

```matlab
%remove the noisy Batch 8 batteries
nind = [3, 40:41];
batch3(nind) = [];
numBat3 = size(batch3,2);
```

```matlab
batch_combined = [batch1, batch2, batch3];
numBat = numBat1 + numBat2 + numBat3;
```

```matlab
%optionally remove the batteries that do not finish in Batch 1; depending
%on the modeling goal, you may not want to do this step
batch_combined([9,11,13,14,23]) = [];
numBat = numBat - 5;
numBat1 = numBat1 - 5;
```

```matlab
batch_combined([119]) = [];
numBat = numBat - 1;
numBat3 = numBat3 - 1;
clearvars -except batch_combined numBat1 numBat2 numBat3 numBat
```

```matlab
%% Output variable
%Extract the number of cycles to 0.88; this is the output variable used in
%modeling for the paper

bat_label = zeros(numBat,1);
for i = 1:numBat
    if batch_combined(i).summary.QDischarge(end) < 0.88
        bat_label(i) = find(batch_combined(i).summary.QDischarge < 0.88,1);

    else
        bat_label(i) = size(batch_combined(i).cycles,2) + 1;
    end
 end
```

```matlab
Y=bat_label;%Response Variable is Cycle Life of cell till it reaches 80% capacity
n=numBat;% Number of training cells
delQ=zeros(n,1);%Discharge capacity difference between cycle 100 and 10
IRdiff=zeros(n,1);%Internal Resistance difference between cycle 2 and 100
Avgct=zeros(n,1);%Mean of the charge time of cycles 2 to 6
```

```matlab
    Tint=zeros(n,1);%Integral of Average Temperature over cycles 2 to 100
    Tmax=zeros(n,1);%Max Temperature from cycle 2 to 100
    Discharge_curve=cell(n,2);%Cell containing matrix of discharge capacity as a function of cycle
    B=cell(n,1);
    Slope_dis=zeros(n,1);%Slope of discharge curve
```

```matlab
%Calculation of Features
for i=1:n
    delQ(i,1)=[batch_combined(i).summary.QDischarge(100,1)-batch_combined(i).summary.QDischarge
    IRdiff(i,1)=[batch_combined(i).summary.IR(100,1)-batch_combined(i).summary.IR(2,1)];
    Avgct(i,1)=[mean(batch_combined(i).summary.chargetime(2:6,1))];
    Tint(i,1)=trapz(batch_combined(i).summary.cycle(2:100,1),batch_combined(i).summary.Tavg(2:1
    Discharge_curve{i,1}=[batch_combined(i).summary.QDischarge(2:100,1)];
    Discharge_curve{i,2}=[batch_combined(i).summary.cycle(2:100,1),ones(99,1)];
    B{i,1}=lasso(Discharge_curve{i,2},Discharge_curve{i,1},"lambda",0);
    Slope_dis(i,1)=B{i,1}(1,1);
    Tmax(i,1)=[max(batch_combined(i).summary.Tmax(2:100,1))];
end
min_delQ=repmat(log10(abs(min(delQ))),n,1);%Minimum discharge capacity
mean_delQ=log10(meanabs(delQ));%Mean of discharge capacity
var_delQ=repmat(log10(abs(sumsqr(delQ-mean_delQ)/(n-1))),n,1);%Variance of discharge capacity
```

```matlab
X=[min_delQ, var_delQ, IRdiff, Avgct, Tmax, Slope_dis, Tint];% 7-Feature matrix
test_battery=[1:1:numBat]';
[Y_sorted, I]=sort(Y);% Sort obeserved Battery cycle life in descending order
```

```matlab
%Gaussian Process Regression
model_rgp=fitrgp(X,Y,"KernelFunction","ardexponential","Beta",1000,"sigma",100);
[Ypred_rgp, Ypred_rgp_std]=predict(model_rgp,X);
Ypred_rgp_sorted=Ypred_rgp(I);% Sorting predicted Battery cycle life
Ypred_rgp_std_sorted=Ypred_rgp_std(I);% Sorting standard deviations
sigmaL = model_rgp.KernelInformation.KernelParameters(1:end-1); % Learned length scales
weights = exp(-sigmaL); % Predictor weights
weights = weights/sum(weights); % Normalized predictor weights
```

```matlab
%Plot results for GPR
%Plot Predictor weights
figure;
semilogx(weights,'ro');
xlabel('Predictor index');
ylabel('Predictor weight');
```

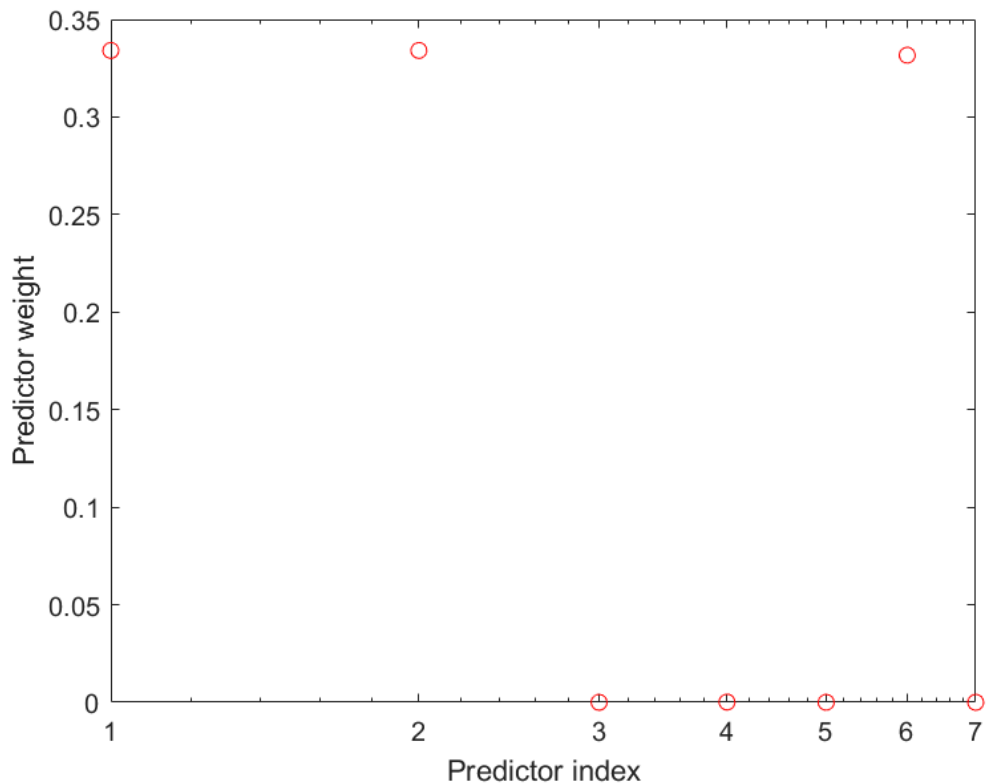

```
%Calculate error values for GPR
[rmse_rgp, per_err_rgp]=error(Y,Ypred_rgp)% RMSE and Error Percentage
```

```
rmse_rgp = 51.7623
per_err_rgp = 4.9613
```

```
Yres_rgp=Y-Ypred_rgp;% Residuals

%Plot Obeserved Cycle Life vs Predicted Cycle Life
plot(test_battery,Y_sorted,".b")
hold on
plot(test_battery,Ypred_rgp_sorted,'r',"Linewidth",1)
plot(test_battery,Ypred_rgp_sorted+Ypred_rgp_std_sorted,'k:')
plot(test_battery,Ypred_rgp_sorted-Ypred_rgp_std_sorted,'k:')
legend("Cycles observed","Cycles predicted","Predicted + Std deviation","Predicted - Std deviat
ylim([0 3000])
xticks(0:10:130)
title("Gaussian Process Regression")
xlabel('Battery')
ylabel('Battery Cycle Life')
hold off
```

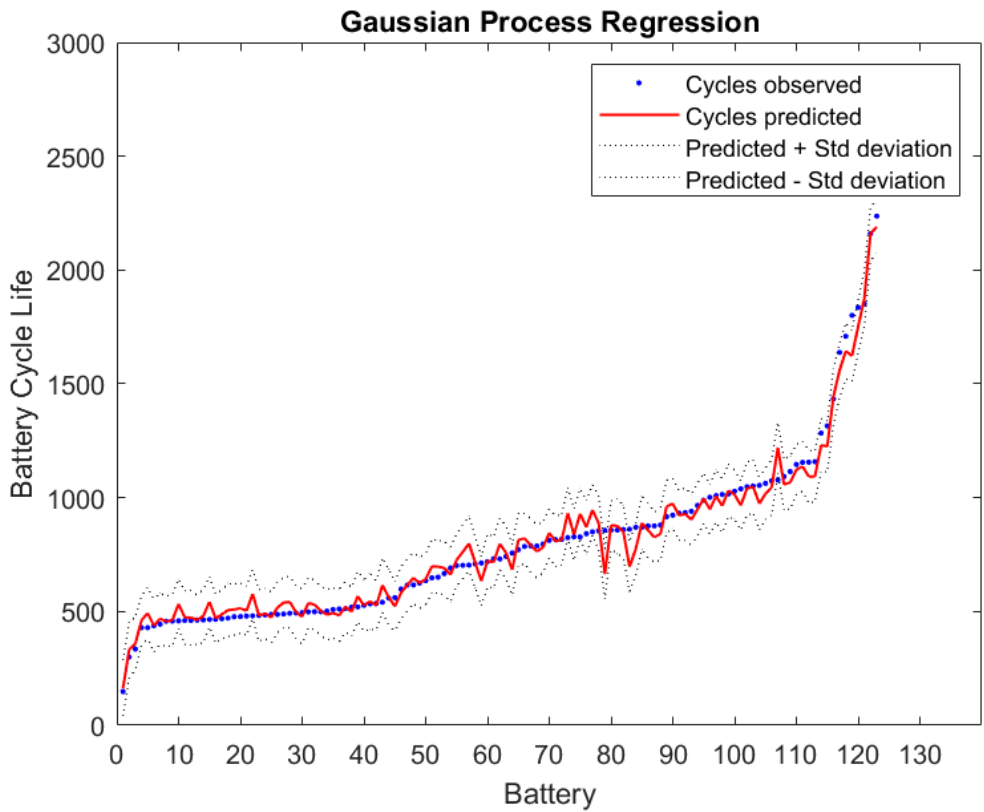

```
%Elastic net
[B_en,FitInfo]=lasso(X,Y,"Alpha",0.5,"CV",4);
idxLambdaMinDeviance = FitInfo.IndexMinMSE;%Find index of coefficients with minimum MSE
B0 = FitInfo.Intercept(idxLambdaMinDeviance);% Interecpt with least MSE
coef = [B_en(:,idxLambdaMinDeviance)];
Ypred_en=B0+X*coef;%Predicted Battery Cycle life
Ypred_en_sorted=Ypred_en(I);% Sorting predicted Battery Cycle life
weights_en =coef; % Predictor weights
weights_en = weights_en/sum(abs(weights_en)); % Normalized predictor weights
```

```
%Plot results for Elastic Net Regression
%Cross-validated MSE for Elastic Net Regression
lassoPlot(B_en,FitInfo,'PlotType','CV');
legend('show') % Show legend
```

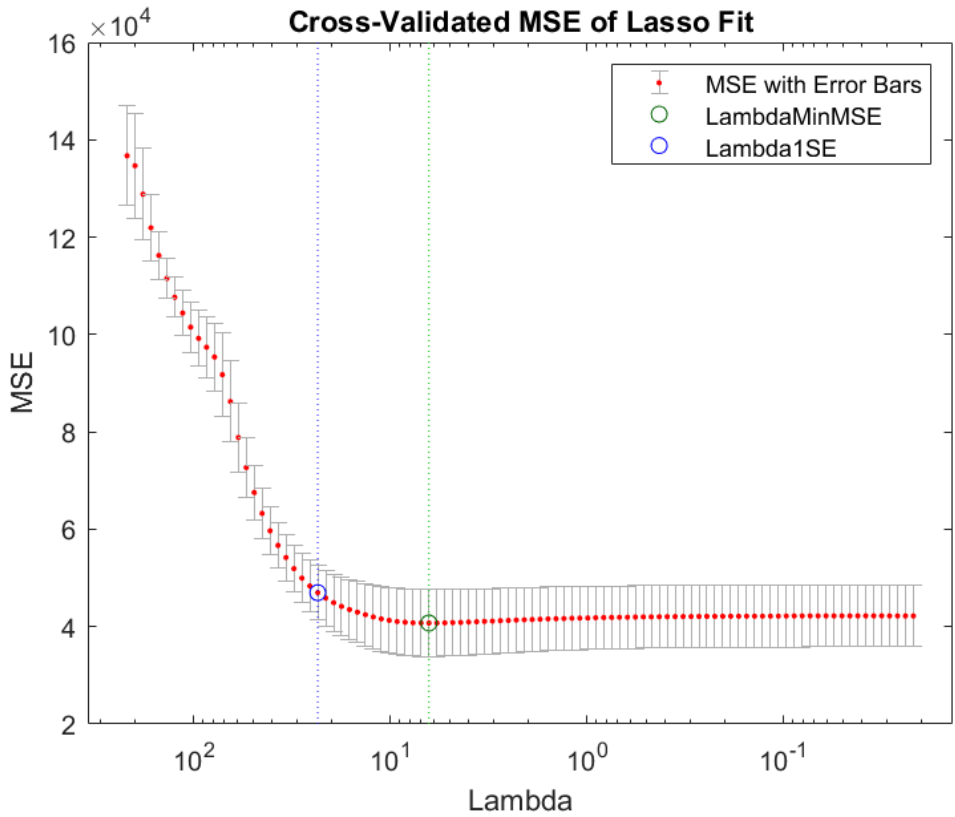

```
%Plot Predictor weights
figure;
semilogx(weights_en,'ro');
xlabel('Predictor index');
ylabel('Predictor weight');
```

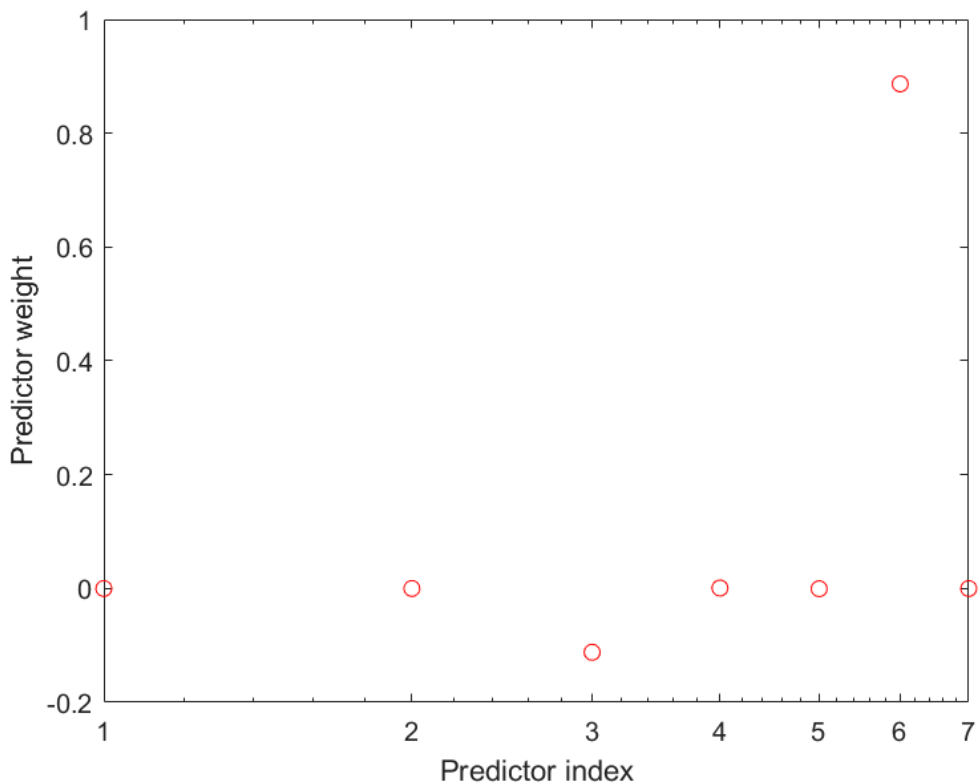

```matlab
%Calculate error  values
[rmse_en, per_err_en]=error(Y,Ypred_en)%RMSE and Error Percentage
```

```
rmse_en = 197.3176
per_err_en = 18.3635
```

```matlab
Yres_en=Y-Ypred_en;%Residuals

%Plot Obeserved Cycle Life vs Predicted Cycle Life
plot(test_battery,Y_sorted,".b")
hold on
plot(test_battery,Ypred_en_sorted,'r',"Linewidth",1)
legend("Cycles observed","Cycles predicted");
ylim([0 3000])
xticks(0:10:130)
title("Elastic Net Regression")
xlabel('Battery')
ylabel('Battery Cycle Life')
hold off
```

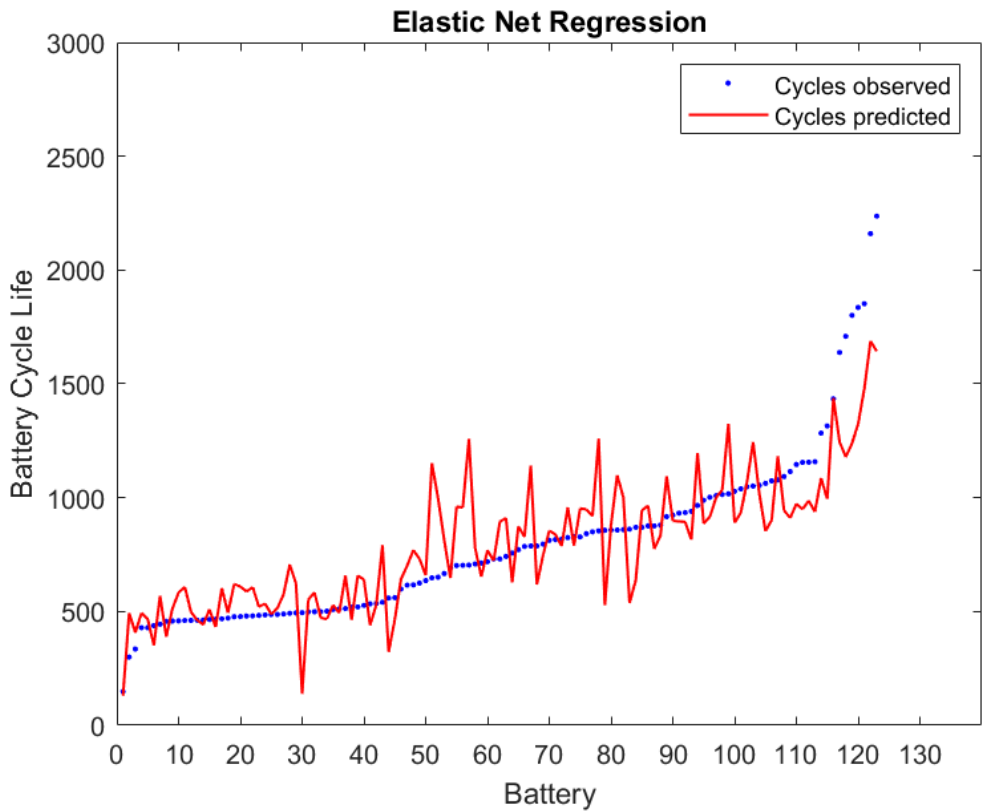

```
figure()
scatter(Y,Ypred_rgp,"*r")
xlabel('Observed Cycle Life')
ylabel('Predicted Cycle Life')
title("Gaussian Process Regression")
refline(1,0)
hold on
yl = get(gca, 'YLim');
plot( [100 100], yl,"k--" )
axes('Position',[0.7 0.2 0.25 0.25])
box on
histogram(Yres_rgp)%GPR Residuals
box off
hold off
```

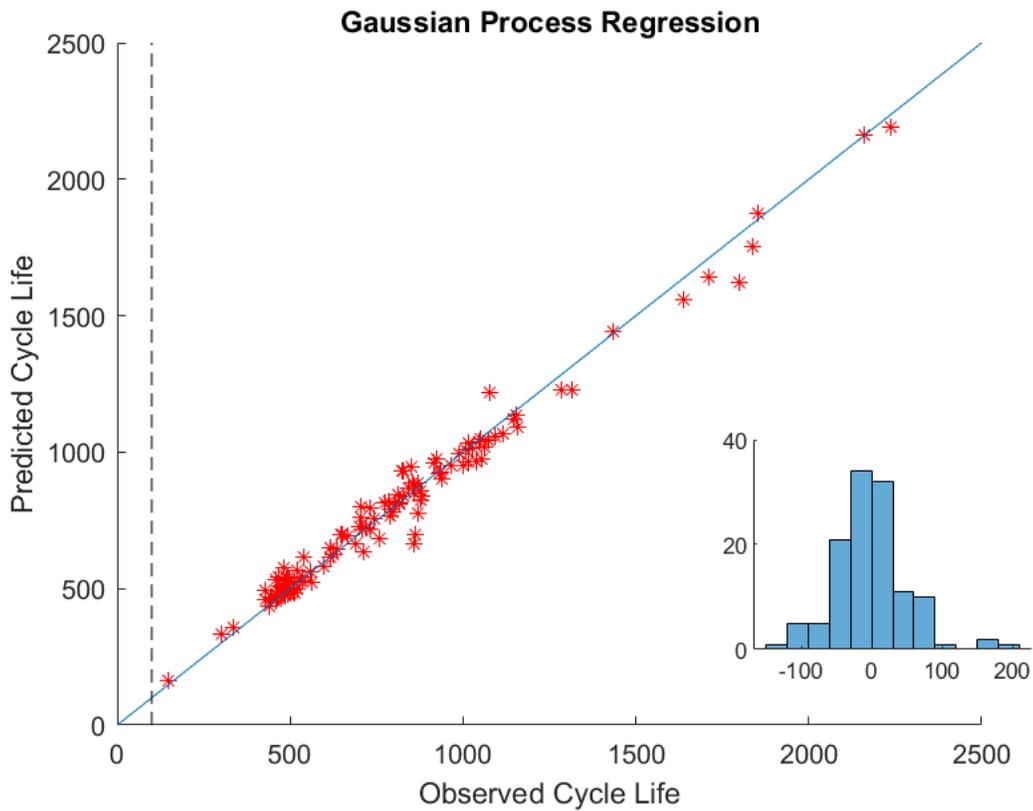

```
figure()
scatter(Y,Ypred_en,"*r")
xlabel('Observed Cycle Life')
ylabel('Predicted Cycle Life')
title("Elastic Net Regression")
refline(1,0)
hold on
yl = get(gca, 'YLim');
plot( [100 100], yl,"k--" )
axes('Position',[0.7 0.2 0.25 0.25])
box on
histogram(Yres_en)%Elastic Net Residuals
box off
hold off
```

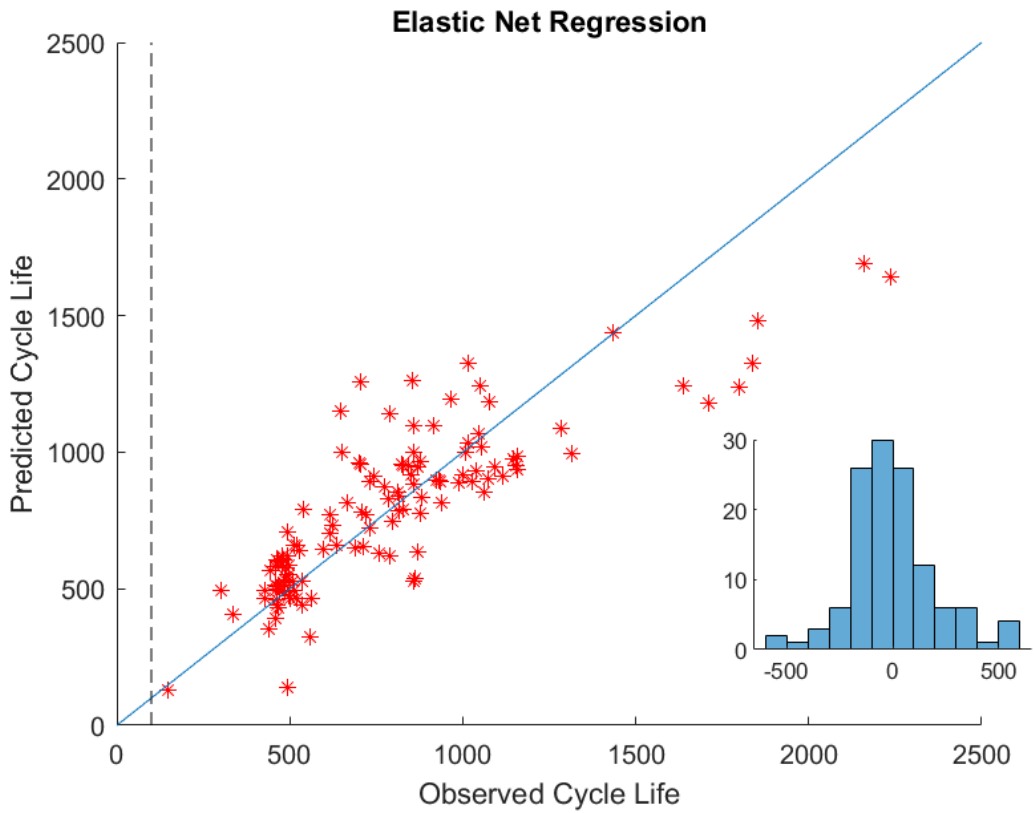

```
%Error calculation function
```

```
function [rmse, per_err]=error(y,ypred)
rmse=sqrt(sum((y(:,1)-ypred(:,1)).^2)/numel(y));
per_err=(sum(abs((y(:,1)-ypred(:,1))./y(:,1))))*100/numel(y);
end
```